\documentclass[twocolumn,pre,showpacs,superscriptaddress,10pt]{revtex4-1}
\usepackage{graphicx}
\usepackage{amsmath}
\usepackage{amsfonts}
\usepackage[table]{xcolor}
\usepackage[caption=false]{subfig}
\usepackage{array}
\usepackage{floatrow}
\usepackage{pbox}
\usepackage[colorlinks=true,linkcolor=blue,citecolor=blue,filecolor=cyan,urlcolor=blue,breaklinks=true]{hyperref}

\begin{document}
\title{Subharmonic oscillations in stochastic systems under periodic driving}

\author{Lukas Oberreiter$^1$, Udo Seifert$^1$, and Andre C. Barato$^2$}
\affiliation{$^1$ II. Institut f\"ur Theoretische Physik, Universit\"at Stuttgart, 70550 Stuttgart, Germany\\
$^2$ Department of Physics, University of Houston, Houston, Texas 77204, USA}

\parskip 1mm
\def\d{{\rm d}}
\def\Ps{{P_{\scriptscriptstyle \hspace{-0.3mm} s}}}
\def\MF{{\mbox{\tiny \rm \hspace{-0.3mm} MF}}}
\def\ts{\tau_{\textrm{sig}}}
\def\tos{\tau_{\textrm{osc}}}
\begin{abstract}
Subharmonic response is a well known phenomena in, e.g., deterministic nonlinear 
dynamical systems. We investigate the conditions under which
such subharmonic oscillations can persist for a long time in open
systems with stochastic dynamics due to thermal fluctuations. In
contrast to stochastic autonomous systems in a stationary state, for which the
number of coherent oscillations is fundamentally bounded by the number of
states in the underlying network, we demonstrate that in periodically
driven systems, subharmonic oscillations can, in principle, remain coherent 
forever, even in networks with a small number of states. 
We also show that, {\sl inter alia}, the thermodynamic cost rises only 
logarithmically with the number of coherent oscillations in a model calculation 
and that the possible periods of the persistent subharmonic response grow linearly
with the number of states. We argue that our results can be relevant for biochemical 
oscillations and for stochastic models of time-crystals.
\end{abstract}

\maketitle
\section{Introduction}

An important class of nonequilibrium systems are those driven time-periodically, which includes 
systems in contact with a heat bath described by stochastic 
dynamics \cite{bran15,raz16,bara16} and closed quantum system with a time-periodic 
Hamiltonian \cite{moes17}. Periodically driven systems display quite rich phenomena that have 
been unveiled recently. Examples range from the realization of 
micro-sized heat engines \cite{schm08,blic12,mart16} for systems in contact with a heat 
bath to many body localization \cite{pont15,laza15,aban16} for closed quantum systems.   
 
A rather basic issue that has not been investigated in detail yet concerns 
the onset of sub-harmonic oscillations, i.e., oscillations of some observable 
with a period larger than the period of the driving in a periodically driven 
system in contact with a heat bath. Does noise necessarily eliminate these oscillations 
in a finite system? For how long can they survive? What is the relation between 
the coherence of these noisy oscillations and energy dissipation? These 
are particularly relevant questions within the following two contexts. 
 
First, temporal oscillations are an essential physical phenomena in living systems. 
Examples include the cell cycle \cite{ferr11}, bacterial circadian rhythms 
\cite{naka05,dong08,john11} and other biochemical oscillations \cite{gold97,nova08}. 
Genetic oscillators have also been designed in synthetic biology \cite{elow00,kim11,potv16}. 
Noise can play an important role in these phenomena, due to the low number of constituents
in a biological system such as a cell. In fact, the effect of noise in such oscillations has 
been extensively investigated \cite{bark00,gonz02,mein02,falc03,mcka07}.    

Most of the analysis of this fundamental issue has been restricted to autonomous systems 
that are not under the influence of periodic driving. In this case, 
the precision of the oscillations can be quantified by the number of coherent oscillations 
\cite{more07,jorg18,qian00,cao15,bara17a}, which is given by the relaxation time divided by 
the period of oscillation. The relation between the precision 
of noisy oscillations in autonomous systems and thermodynamics has received much 
attention recently \cite{qian00,cao15,bara17a,nguy18,fei18,wier18,mars19}. 

However, biophysical oscillations often take place under the influence of 
a periodic signal. In fact,  a striking feature 
about biological oscillations is the range of periods involved \cite{gold08}. For instance, 
if we consider oscillations within a cell, calcium oscillations \cite{falc04} can have 
a period of the order of minutes, whereas circadian clocks have a period of a day. 
Therefore, generalizing this example, a relevant question is whether a fast periodic signal 
influences the precision of oscillations that happen at a slower time-scale. In other words,
can a fast signal with, say, a period of the order of seconds or minutes lead to precise 
oscillations with a period of a day?   

Second, a main feature of so called time-crystals \cite{sach15,khem16,else16} is a subharmonic response
of an observable. These are many-body closed quantum systems that also display 
spatial long-range order as a main feature. Experimental realizations of time-crystals 
have been performed in \cite{zhan17,choi17}. Time-crystals can also be observed in 
dissipative open systems in contact with a heat bath \cite{laza17,yao18,gong18,wang18,gamb19}. 
Hence, the fundamental limits of the effect of thermal noise on subharmonic oscillations can 
become relevant for such time-crystals as well.

The general purpose of this paper is to provide a first answer to the questions 
raised above about subharmonic oscillations in a stochastic systems under 
the influence of periodic driving. Our main result is that even in a finite system
with  thermal fluctuations, sub-harmonic oscillations can survive for 
an arbitrarily long time, with no fundamental limit. We prove this result by
introducing a simple model that exhibits an indefinite number of coherent subharmonic oscillations in a 
particular limit. Such a divergent number of coherent oscillations in {\it periodically driven} systems is 
in contrast with the oscillatory behavior in {\it autonomous} systems, for which 
the number of coherent oscillation is fundamentally limited by the number
of states of the underlying network \cite{bara17a}. 

We also show that the trade-off between energy dissipation and the number of coherent 
oscillations is always more advantageous for our periodically driven model  as compared to
an autonomous system, if the number of coherent oscillations is large. 
Finally, we discuss a mathematical condition on the so-called fundamental matrix from Floquet 
theory  that leads to restrictions on the possible number of coherent subharmonic 
oscillations that can be achieved in a periodically driven system.

\section{General setup}

We consider a Markov process with $\Omega$ discrete states and continuous time $t$. 
The transition rate from state $i$ to state $j$ at time $t$ is denoted by 
$w_{ij}(t)$. These transition rates are time-periodic with a period $\ts$, 
i.e., $w_{ij}(t+\ts)=w_{ij}(t)$. The time evolution of the probability to 
be in state $i$ at time $t$, which is denoted $p_i(t)$, follows the master 
equation 
\begin{equation}
\frac{d}{dt}\mathbf{p}(t)=\mathcal{L}(t)\mathbf{p}(t),
\end{equation}
where $\mathbf{p}(t)$ is a vector with components $p_i(t)$ and 
$[\mathcal{L}(t)]_{ji}\equiv (1-\delta_{ij})w_{ij}(t)-\delta_{ij}\sum_k w_{ik}(t)$ 
are the elements of the stochastic generator $\mathcal{L}(t)$. 


The solution of this equation is given by
\begin{equation}
\mathbf{p}(t)= \sum_{a=1}^{\Omega} C_a\textrm{e}^{\mu_a t}\mathbf{p}^a(t), 
\label{eqcorrgen}
\end{equation}
where $C_a$ are constants that depend on the initial condition, 
$\mathbf{p}^a(t)=\mathbf{p}^a(t+\ts)$, and  $\mu_a$ are the so-called 
Floquet exponents \cite{klau08}. The Floquet exponent with largest 
real part is $\mu_1=0$. In the long time limit, the probability 
converges to the eigenvector $\mathbf{p}^1(t)$. 

Sub-harmonic oscillations with a period $\tos$ are quantified by the 
Floquet exponent with the second largest real part $\mu_2= -X_R\pm X_Ii$,
where $\tos=2\pi/X_I$ and $X_R$ is the inverse of the relaxation 
time. We are interested in the ratio 
\begin{equation}
\mathcal{R}\equiv X_I/X_R.
\label{defR}
\end{equation}  
The number of coherent oscillations  is given by $\mathcal{R}/2\pi$. This ratio 
then quantifies the precision of  subharmonic oscillations.

Floquet exponents can be calculated from the fundamental matrix \cite{grim90,klau08} 
\begin{equation}
\mathcal{M}\equiv\overleftarrow{\exp}\left(\int_0^{\ts}\mathcal{L}(t)dt\right),
\label{eqfundmat}
\end{equation}
where the arrow indicates a time-ordered exponential.
The eigenvalues of this matrix $\rho_a$, also known as Floquet multipliers, 
are related to the Floquet exponents as
\begin{equation}
 \mu_a=\ts^{-1}\ln \rho_a.
\label{defmulti}
\end{equation} 
The largest eigenvalue is $\rho_1=1$. The eigenvalue with the second largest real part 
is written as $\rho_2=r\textrm{e}^{i\theta}$, where $r\le 1$. From Eq. \eqref{defmulti}, 
we then obtain $X_R= -\ln r/\ts$ and $X_I= \theta/\ts$. This 
last equality leads to $\tos=2\pi/X_I= 2\pi\ts/\theta\ge \ts$, which shows that the 
period of oscillation is always larger than the period of driving.  

\begin{figure}
\includegraphics[width=77mm]{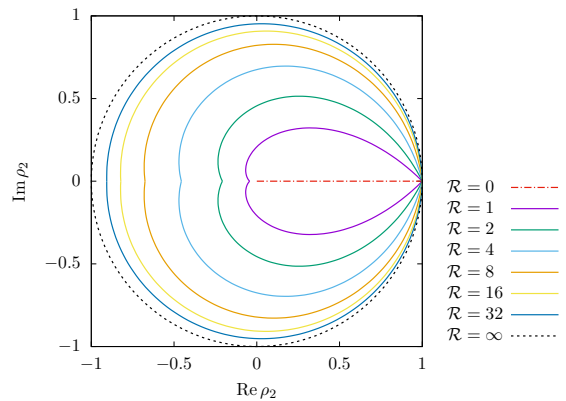}
\vspace{-2mm}
\caption{Contour lines  for constant $\mathcal{R}$ in the complex plane of the eigenvalue $\rho_2$.}
\label{fig1} 
\end{figure}

The contour lines for constant $\mathcal{R}$, which is defined in Eq. \eqref{defR},  in
the complex plane of the eigenvalue $\rho_2$  are shown in Fig. \ref{fig1}. On the circle 
corresponding to $r=1$, the relaxation time $X_R^{-1}$ diverges ($\mathcal{R}\to\infty$) and 
subharmonic oscillations do not decay. We now provide a model with a finite number 
of states for which the eigenvalue $\rho_2$ lies on this circle in a certain limit. 

\section{Limit of indefinite coherent subharmonic oscillations}

\begin{figure}
\includegraphics[width=78mm]{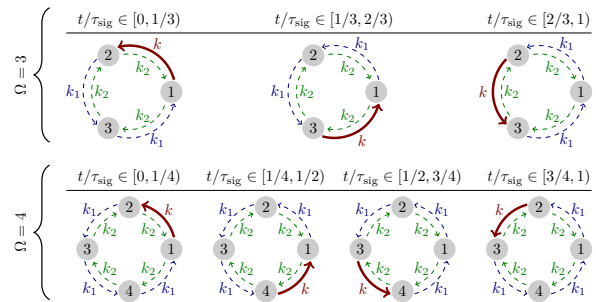}
\vspace{-2mm}
\caption{Illustration of the protocol for $\Omega=3$ and $\Omega=4$. The transition rates 
are $k_1\equiv k/c$ and $k_2\equiv k/c^{(\Omega-1)/\Omega}$.}
\label{fig2} 
\end{figure}

Our model corresponds to a particle on a ring with $\Omega$ discrete states, as illustrated 
in Fig. \ref{fig2}. It can be interpreted as a colloidal particle 
on a ring subject to thermal noise and driven by a periodic protocol. 
The transition rate from state $i$ to state $i+1$ is denoted 
$w_i^+(t)$ and the reversed transition from state $i+1$ to state $i$ is 
denoted $w_i^-(t)$. For $i=\Omega$, one of the next neighbors is $i+1=1$. The period 
of the external signal is divided into $\Omega$ time intervals. In a generic 
time-interval $t\in[(n-1)\ts/\Omega,n\ts/\Omega]$, where $n=1,2,\ldots,\Omega$, there are three different transition 
rates: the reversed transition rates are given by $w_i^-=k/c^{(\Omega-1)/\Omega}$ for all states,
$w_i^+= k$ for state $i=(\Omega+1-n)\mod \Omega+1$, and $w_i^+=k/c$ for all other states with 
$i\neq(\Omega+1-n)\mod \Omega+1$.

These transition rates fulfill detailed balance for fixed time $t$, i.e., 
\begin{equation}
\prod_{i=1}^{\Omega} [w_i^+(t)/w_i^-(t)]=1.
\end{equation}
In terms of energies and energy barriers the transition rates can be written as
$w_i^+(t)=k\textrm{e}^{E_i(t)-B_i(t)}$ and $w_i^-(t)=k\textrm{e}^{E_{i+1}(t)-B_i(t)}$,
where $E_i(t)$ is the energy of state $i$, $B_i(t)$ is the energy barrier between states 
$i$ and $i+1$, and we set temperature $T=1$ and Boltzmann's constant $k_B=1$ throughout.
For the protocol explained in the previous paragraph, for
$t\in[(n-1)\tau/\Omega,n\tau/\Omega]$, the energy difference between two neighbors 
is $E_{i}-E_{i+1}=[(\Omega-1)/\Omega]\ln c$ for $i=(\Omega+1-n)\mod \Omega+1$ and,  
$E_{i}-E_{i+1}=-(1/\Omega)\ln c$ for $i\neq(\Omega+1-n)\mod \Omega+1$.

As a main result, we obtain that this finite stochastic system  
exhibits indefinite subharmonic oscillations in the following particular limit. First, 
with $k\gg \ts^{-1}$, the one internal transition with rate $k$ is much faster than the external signal. 
Second, $c$ is large, which leads to diverging energy differences between neighboring states, in such a way that the other transition
rates are slow in comparison to the external signal, i.e., $k/c\ll \ts^{-1}$ and $k/c^{(\Omega-1)/\Omega}\ll \ts^{-1}$. 
In this limit, by numerical evaluation of the Floquet multipliers we observe that 
$\rho_2= \textrm{e}^{2\pi i/(\Omega-1)}$, which implies indefinite subharmonic 
oscillation with a period $\tos=(\Omega-1)\ts$. As discussed in Appendix \ref{appa}, it is also 
possible to obtain indefinite subharmonic oscillations with a period $\tos=m\ts$,
where $m= 2,3,\ldots,\Omega-1$, by making simple changes to this protocol. Hence, 
the number of possible periods grows linearly with the number of states $\Omega$.

The onset of such indefinite subharmonic oscillations in this particular limit can be understood if we 
consider the case $\Omega=3$ in Fig. \ref{fig2} (a similar explanation holds for general $\Omega$). 
Let us assume that before the first period the particle 
starts at state $1$. During the first part of the period the particle will  jump to state $2$.
During the second part of the period the particle will remain trapped in state $2$ since the transition rates 
to leave this state are slow in comparison to the external signal. During the third part of the 
period the particle jumps from state $2$ to state $3$. Hence, at the end of the first period the particle will 
be in state $3$. For the next period the particle starts at state $3$ and jumps during the middle interval to state $1$, 
where it stays trapped until the end of the period. Therefore, subharmonic oscillations with a period 
$\tos=2\ts$  take place, with the particle at state $3$ at the end of odd
periods and at state $1$ at the end of even periods. 

In this peculiar limit, there is a perfect coupling between the 
dynamics of the system and the external signal that is deterministic, 
i.e., the system gets stuck in a state during each part of the 
period and transitions between states can only happen with a change of
the protocol in the next part of the period. Clearly, this 
deterministic coupling is a general sufficient condition for the onset
of subharmonic oscillations. For more complex models, we can imagine an
effectively similar piecewise protocol that traps the system in some 
region of phase space in each part of the protocol. For such complex 
models, an observable that takes different values for these different 
regions of phase space will then display indefinite subharmonic 
oscillations. Hence, our model serves as a general guiding principle 
to obtain indefinite coherent subharmonic oscillations.

\section{Relation between number of coherent oscillations and energy dissipation}

\begin{figure}
\includegraphics[width=72mm]{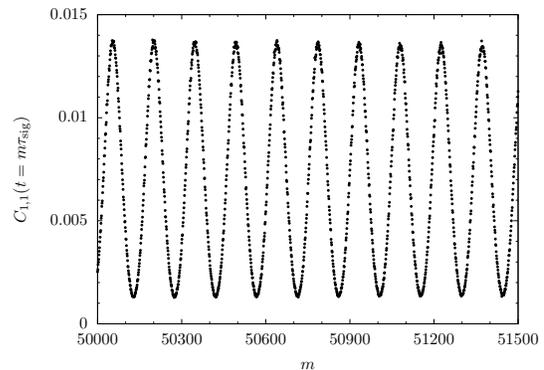}
\vspace{-2mm}
\caption{Stroboscopic plot of the correlation function $C_{1,1}(t)$, which is the 
the probability to be in state $1$ at time $t$ given that the system started at 
state $1$. Parameters are set to $\Omega=145$, $k=600$, $\ts=1$, and $c=10^6$.
The result from the numerics is $\tos\simeq 146\ts$
and $R\simeq 2347$. In the discussion given in the main text $\ts=1$ is set to $10$ minutes, which implies 
$k= 1 s^{-1}$, and thus leads to a period of subharmonic oscillations close to one day.}
\label{fig3} 
\end{figure}

We now compare this periodically driven system to an autonomous system, modeled as   
a Markov process with time-independent transition rates, showing coherent oscillations.  
Two factors that limit the number of coherent oscillations in such autonomous system are 
the number of states and energy dissipation \cite{bara17a}. In an autonomous system with $\Omega$ states, 
$\mathcal{R}\le \cot(\pi/\Omega)$, where $\mathcal{R}$ for an autonomous system is defined as the ratio of imaginary and real parts
of the second largest eigenvalue associated with the Markov generator.
In Fig. \ref{fig3}, we illustrate the fact that 
a fast periodic signal can increase the precision of oscillations on a slower time-scale beyond the limits that would be achievable in 
the absence of the signal. The period of the signal is set to
$\ts= 10$ minutes, from numerical evaluation we obtain that  the period of subharmonic oscillations is $\tos\simeq 1$ day and 
$\mathcal{R}\simeq 2347$. This number is much larger than the fundamental limit for 
an autonomous system with $\Omega=145$ states, which is $\mathcal{R}=\cot(\pi/145)\simeq 46$. 

\begin{figure}
\includegraphics[width=77mm]{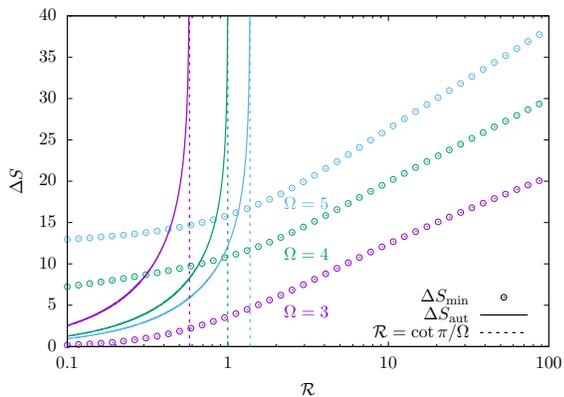}
\vspace{-2mm}
\caption{Entropy change per period of oscillation as a function of $\mathcal{R}$ for an autonomous system 
$\Delta S_{\textrm{aut}}$ and 
a periodically driven system $\Delta S_{\textrm{min}}$, where $\Delta S_{\textrm{min}}$ is 
obtained with a numerical minimization over the parameters $c$ and $k$. The vertical 
dashed lines represent the bound for autonomous systems $\cot(\pi/\Omega)$.}
\label{fig4} 
\end{figure}

Let us now analyze the thermodynamic cost. In the limit where the model achieves indefinite subharmonic oscillations, energy differences have to be large, i.e., $\ln c\gg 1$. 
This condition implies that  the  entropy production that characterizes the thermodynamic cost diverges. The expression for the rate of 
entropy production \cite{seif12}
\begin{equation}
\sigma= \ts^{-1}\int_0^{\ts}dt\sum_{i,j}p_i^1(t)w_{ij}(t)\ln(w_{ij}(t)/w_{ji}(t)),
\end{equation}
where $p^1_i(t)=p^1_i(t+\ts)$ is the time-periodic distribution, leads to 
the amount of entropy change in one period of subharmonic oscillation as $\Delta S\equiv\sigma\tos$.

For the comparison of the trade-off between thermodynamic cost and $\mathcal{R}$ between our model and an autonomous system, 
we consider the following model for an autonomous system. A particle jumps on a 
ring with $\Omega$ states. The time-independent transition rates are given by $w_i^+=k\textrm{e}^{F/\Omega}$ and $w_i^-=k$, 
where $F=\ln\prod_{i=1}^{\Omega} [w_i^+/w_i^-]$ is the force that drives the autonomous system out 
of equilibrium. This particular choice of uniform rates $w_i^+$ and $w_i^-$ maximizes $\mathcal{R}$ for a fixed force $F$  \cite{bara17a}.
For this model $\mathcal{R}=\cot(\pi/\Omega)\tanh(F/(2\Omega))$ and the thermodynamic cost of 
one period of oscillation is $\Delta S_{\textrm{aut}}\equiv \sigma\tos= 2\pi F/[\Omega\sin(2\pi/\Omega)]$ \cite{bara17a}. 

In Fig. \ref{fig4}, we compare the thermodynamic cost per period that is required to obtain a certain number of 
coherent oscillations for both the periodically driven system and the autonomous system. For the periodically 
driven system, we have performed a numerical minimization of $\Delta S$ with the 
constraint that $\mathcal{R}$ is fixed. The free parameters in this minimization are $k$ and $c$. As expected, the 
amount of required thermodynamic cost increases with $\mathcal{R}$. In particular, for the periodically driven system 
the thermodynamic cost increases logarithmic with $\mathcal{R}$. For large enough 
$\mathcal{R}$ the required cost for the periodically driven system is smaller than the one
for the autonomous system. This property is a consequence of the fact that the autonomous system is bounded 
by $\mathcal{R}\le \cot(\pi/\Omega)$ and this bound is reached at a limit of diverging thermodynamic cost, 
whereas the periodically driven system is not constrained by this bound. 
      
\section{Generic constraints}

\begin{figure}
\includegraphics[width=77mm]{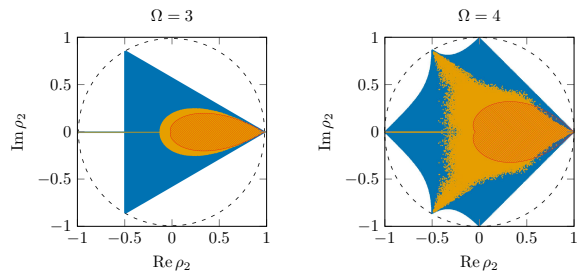}
\vspace{-2mm}
\caption{Domain of the the possible eigenvalues $\rho_2$ for $\Omega=3$ and $\Omega=4$.
The larger blue region represents the constraint on stochastic matrices obtained in \cite{dmit45,dmit46}.
The yellow scatter plots were generated with random matrices that fulfill the constraint in Eq. 
\eqref{eqrest}. The region inside the smaller red line represents $\mathcal{R}\le \cot(\pi/\Omega)$.
}
\label{fig5} 
\end{figure}

We finally discuss  generic, model independent constraints on sub-harmonic oscillations. 
The mathematical properties of the fundamental matrix $\mathcal{M}$ in Eq. \eqref{eqfundmat} restrict
 the accessible region in the complex plane of the eigenvalue 
$\rho_2$. First this matrix is a stochastic matrix, i.e., all elements are positive and the sum of the elements
in a column is $1$. Dmitriev and Dynkin \cite{dmit45,dmit46,swif72} have shown that the eigenvalue $\rho_2$ of such a stochastic matrix is 
constrained to lie in the regions plotted in Fig. \ref{fig5}  for $\Omega=3$ and $\Omega=4$.

Second, the fundamental matrix $\mathcal{M}$ is an embeddable matrix \cite{good70,davi10}, which means that it must fulfill the 
constraint  
\begin{equation}
0\le\textrm{det}\mathcal{M}\le \prod_{i=1}^\Omega M_{ii}.
\label{eqrest}
\end{equation}
To study the effects of Eq. \eqref{eqrest} on the eigenvalue $\rho_2$ we have 
generated random matrices that fulfill this constraint and calculated their eigenvalues 
numerically for $\Omega=3$ and $\Omega=4$. The results indicate that this 
condition lead to a region smaller than the region for general stochastic matrices, as shown 
in Fig. \ref{fig5}. In particular, the accessible region on the circle with $r=1$, where $\rho_2=r\textrm{e}^{i\theta}$, 
are the points $\rho= \textrm{e}^{2\pi i/m}$, where $m=2,3, \ldots, \Omega-1$. We have confirmed 
this result up to $\Omega=6$. This numerics suggests that for a stochastic system with $\Omega$ 
states and an indefinite number of sub-harmonic oscillations, the period of oscillation is 
given by  $\tos=m\ts$, where $m=2,3, \ldots, \Omega-1$. Indeed these are the periods of 
subharmonic oscillations that we have obtained in our model. For an autonomous system 
corresponding to a continuous time Markov process with constant 
transition rates, we have $\mathcal{R}\le \cot(\pi/\Omega)$ \cite{bara17a}. This restriction leads to 
the smaller regions shown in Fig. \ref{fig5}.

\section{Conclusion}

In summary, we have shown that, in principle, subharmonic oscillations in a finite, periodically driven system
subjected to thermal noise can survive for an arbitrarily long time. 
In contrast to autonomous systems, the number of states does not 
impose any fundamental constraint on the maximal number of coherent oscillations.  
We have analyzed the trade-off relation between the number of coherent oscillations 
and energy dissipation for our model, showing that an increase in energy dissipation 
can lead to a larger number of coherent oscillations and that a periodically driven 
system can achieve the same number of coherent oscillations with a smaller energy 
budget compared to an autonomous system.

Our work raises a few fundamental questions related to the relation between 
thermodynamics and subharmonic oscillations in periodically driven systems. Is 
a diverging rate of energy dissipation that we found in our model a necessary condition 
to achieve the formal limit of indefinite subharmonic oscillations? What are 
the optimal protocols that minimize energy dissipation for a given number of 
coherent oscillations? What is the minimal energy budget for a given number 
of coherent oscillations? Furthermore, analyzing the thermodynamics 
of stochastic models for time-crystals, such as the one introduced in \cite{yao18}, is an 
interesting direction for future work. The deterministic coupling leading to indefinite 
oscillations introduced here can be helpful to build such models.

Concerning biochemical oscillations, we have introduced the idea that a fast 
periodic signal can dramatically improve the coherence of  oscillations with a 
longer period. Understanding under which conditions this feature takes place in 
more realistic and complex models for biochemical oscillations under the influence of
such a signal is an appealing future step.  This idea could, in principle, be used to 
build precise synthetic biochemical oscillators in an experimental setup where 
the system is under the influence of a periodic signal. Furthermore, an experimental 
verification of a large number of coherent subharmonic oscillations in a periodically driven
finite system with thermal fluctuations could be realized with laser-induced modulated 
energy landscapes for a driven colloidal particle, see e.g. \cite{curt02,blic07,mart16a,loza16,Gavr17}, 
by mimicking the simple model we have used as a proof of concept.

\begin{figure*}[h]
\includegraphics[width=1.\textwidth]{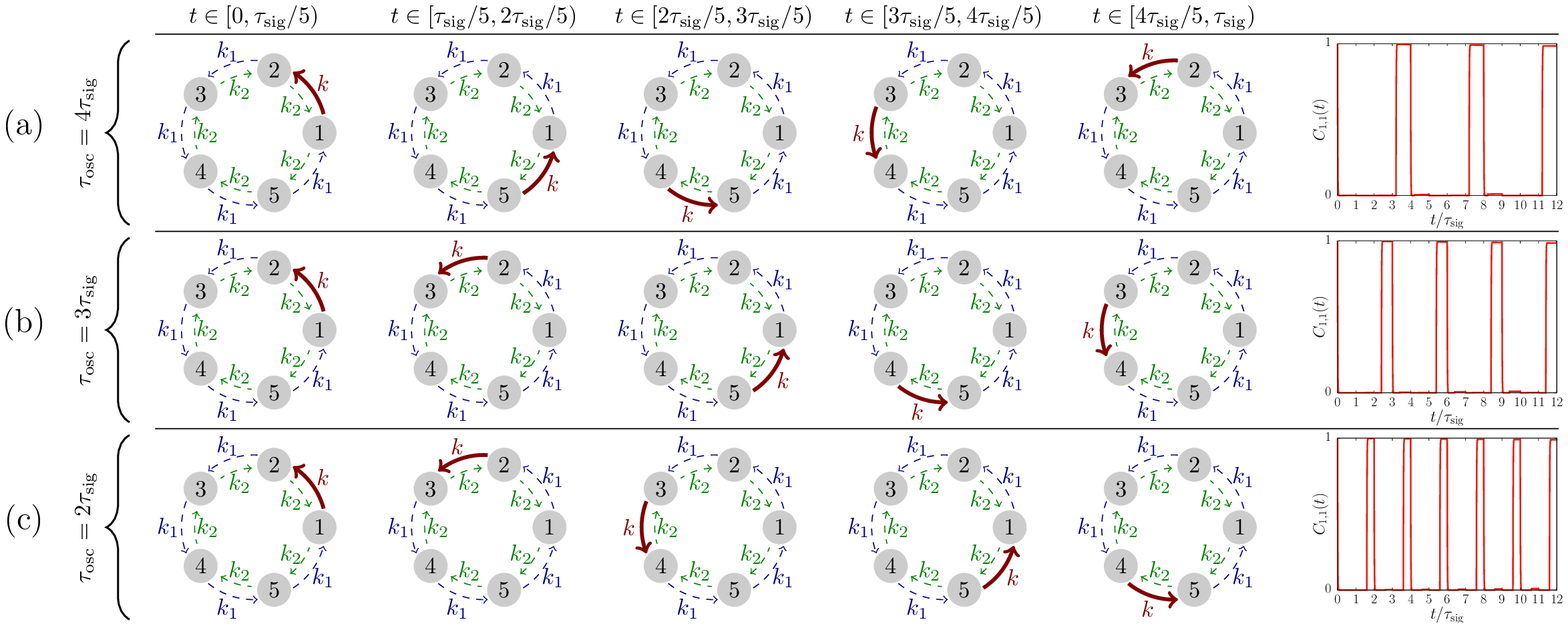}
\caption{Three different protocols for $\Omega=5$ states. (a) Protocol in the main text with $\tau_\mathrm{osc}=4\tau_\mathrm{sig}$.
(b) Protocol with $\nu=1$ anticlockwise steps, which leads to $\tau_\mathrm{osc}=3\tau_\mathrm{sig}$. 
(c) Protocol with $\nu=2$ anticlockwise steps, which leads to $\tau_\mathrm{osc}=2\tau_\mathrm{sig}$.
On the right, the respective correlation function $C_{1,1}(t)$, which is the probability to be in state $1$ given that initially the particle was in state $1$, obtained 
from numerical simulations for $k=100$, $k_1=10^{-4}$, $k_2\approx 1.6\cdot 10^{-3}$ and $\tau_\mathrm{sig}=1$.}
\label{fig:prot_5}
\end{figure*}

\appendix

\section{Protocols for indefinite subharmonic oscillations with different periods}
\label{appa}

For the specific protocol discussed in the main text, which is sketched in the upper panel of Fig.~ \ref{fig:prot_5} for $\Omega=5$ states, an indefinite number of subharmonic oscillations sets in for 
the limit $k_1, k_2 \ll \tau_\mathrm{sig}^{-1} \ll k$. The period of oscillations is $\tau_\mathrm{osc}=(\Omega-1)\tau_\mathrm{sig}$. For $\Omega=5$ we can understand this result 
in the following way. Consider a particle initially in state $1$ in the upper panel of Fig.~(\ref{fig:prot_5}). For this particular limit, after the first period the particle will be in 
state $3$. In the susequent periods the particle will move one position in the anti-clockwise direction per period, going to state $4$ after the second period and to state $5$ after the
third period. After the fourth period the particle will be back at state $1$. Hence, the model displays subharmonic oscillations with  $\tau_\mathrm{osc}=4\tau_\mathrm{sig}$.

Guided by this simple dynamics, we construct protocols, which have periods of subharmonic oscillations given by $\tau_\mathrm{osc} = m \tau_\mathrm{sig}$ with $m = 2,3,\dots,\Omega-1$. 
They are obtained by the following modifications in relation to the protocol in the main text.
In the standard protocol the transition associated with the strongest rate $k$ moves one clockwise position after each protocol step.
In the modified protocol, the rate $k$ moves one anti-clockwise position for each of the first $\nu$ steps. Then, in the next step, this rate $k$ moves $\nu+1$ positions in the clockwise direction, 
going to the transition from state $\Omega$ to state $1$. In the subsequent steps, the transition rate $k$ moves one position in the clockwise direction. 
For $\Omega=5$, these modified protocols with $\nu=1$ and $\nu=2$ are shown in Fig.~\ref{fig:prot_5}. From this figure one 
can see that the period of oscillation for a protocol with $\nu$ anti-clockwise steps is  $\tau_\mathrm{osc} = (\Omega-1-\nu) \tau_\mathrm{sig}$.

\bibliographystyle{apsrev4-1}

\bibliography{refs} 

\end{document}